\documentclass[manuscript,preprint]{aastex}

\usepackage{graphicx}
\usepackage{amsmath}

\usepackage{epsfig}
\usepackage[dvips]{color}

\usepackage{subfigure}
\usepackage{longtable}\usepackage{lscape}
\usepackage{times}
\DeclareGraphicsExtensions{.pdf, .jpg}

\usepackage{color,soul}

\shorttitle{Bondi-Hoyle Accretion around EGB Black Hole}
\shortauthors{O.Donmez}

\begin{document}

\title{Bondi-Hoyle Accretion
  around the Non-rotating Black Hole in 4D Einstein-Gauss-Bonnet Gravity}

\author{O. Donmez\altaffilmark{1}}
\affil{College of Engineering and Technology, American
  University of the Middle East (AUM), Kuwait}

\altaffiltext{1}{College of Engineering and Technology, American
  University of the Middle East (AUM), Kuwait}




\begin{abstract}

  In this paper, the numerical investigation of a Bondi-Hoyle accretion around a
  non-rotating black hole in a novel four dimensional Einstein-Gauss-Bonnet
  gravity is investigated by solving the general relativistic hydrodynamical equations
  using the high resolution shock capturing scheme. For this purpose, the accreated
  matter from the wind-accreating $X-$ray binaries falls towards the black hole
  from the far upstream side  of the domain, suphersonically. We study the effects of
  Gauss-Bonnet coupling constant $\alpha$  in $4D$ EGB gravity on the accreated matter
  and shock cones created in the downstream region in detail.
  The required time having the shock cone in downstream region is getting smaller
  for $\alpha > 0$ while it is increasing for  $\alpha < 0$.
  It is found that 
  increases in  $\alpha$ leads violent oscillations inside the shock cone and  increases
  the accretion efficiency. The violent oscillations
  would cause increase in the energy flux, temperature, and spectrum of $X-$ rays. So
  the quasi-periodic oscillations (QPOs) are naturally produced inside the shock cone
  when  $-5 \leq \alpha \leq 0.8$.
  It is also confirmed that EGB black hole solution converges to the Schwarzschild one in
  general relativity when $\alpha \rightarrow 0$.
  Besides, the negative coupling constants also give reasonable physical solutions and
  increase of $\alpha$ in negative directions suppresses the possible oscillation
  observed in the shock cone.
  
\end{abstract}



\section{Introduction}
\label{Introduction}

Einstein's general theory of relativity has successfully passed many
observational tests during the last decades
\citep{Akiyama1, Akiyama2, Akiyama3, Abbott1, Abbott2}.
The existence of super-massive
black hole at the center of galaxy M87 was one of the most impressive discoveries.
The observed M87 black hole shadow indicates that the observed data well
consistent with the prediction of the general theory of relativity.
The observed gravitational waves created during the coalescing of
two stellar mass black holes made by LIGO detectors was another discovery
to see the ripples of the space time predicted by Einstein's general theory
of relativity. The observed electromagnetic spectrum emitted from
an accretion disk around the black holes could allow us to define many
properties of the black holes, such as mass, spin, etc.\citep{Frank1, Yuan1, Nampalliwar1}.
All recent observations indicate that testing the general theory of
 relativity and its alternatives in a week and a strong gravitational
regions are hot topics in this century. On this end, in order to extract more
detailed information about the strong gravitational regimes around the black holes, we 
study the alternative theory to the Einstein's theory of the relativity
called four dimensional Einstein-Gauss-Bonnet (EGB) gravity which was
first formulated by \citet{Glavan1}.

A novel four-dimensional EGB gravity theory
was proposed in \citet{Glavan1} where the Gauss-Bonnet  coupling constant is 
$\alpha \rightarrow \alpha/(D-4)$ in the limit $D \rightarrow 4$ and bypasses
the Lovelocks theorem. As a consequence of this limit, the Gauss-Bonnet
coupling constant  gives non-trivial contributions to the gravitational dynamics.
It is also shown that the theory is free from Ostrogradsky instability and preserves
the number of degrees of freedom. And the static spherically symmetric black hole,
which has two horizons instead of one compared to the Schwarzschild black hole,
was discovered\citep{Cheng1}. The horizon is rescaled with
the Gauss-Bonnet coupling constant $\alpha$ and the causal structure of the black hole
radically is altered with a repulsive effect. Although this
$4D$ EGB theory is currently under debate \citep{Julio1, Gurses1, Ai1},
it is worthy  and meaningful to
study the spherical symmetric black hole solution of $4D$ EGB gravity.

The newly discovered four-dimensional static and spherically symmetric Gauss-Bonnet
black hole was used to reveal many interesting features of some astrophysical problems.
Lots of work have been done the different astrophysical phenomena the viability  of the
solution. 
The properties of the black hole and its shadow \citep{Konoplya1, Roy1}, the innermost
stable circular orbit  for massive particles and photons \citep{Guo1, Zhang1},
generating and radiating black hole \citep{Ghosh1, Ghosh2},  week cosmic censorship
conjecture \citep{Yang1}, the gravitational lensing
by black holes \citep{Jin1, Islam1}, observational constraints on the Gauss-Bonnet
constant \citep{Feng1, Clifton1},  the  growth  rate  of  non-relativistic  matter
perturbations \citep{Zahra},  and  Greybody factor and power spectra of the
Hawking radiation \citep{Konoplya1, Zhang2} were studied with all details.

The wind accretion scenario onto the black hole is one of the important physical
phenomena to explain soft and hard $X-$ rays observed by different $X-$ ray
telescopes.  The wind accretion process can be explained by using the
Bondi-Hoyle model \citep{Edgar1}. A black hole moving inside the gas cloud
captures mass which causes either acceleration or deceleration of it. As a result of this the shock
cone or bow shock would be formed around the black hole. This is the one of hot topics
studied by different researchers using the analytical and the numerical techniques in the last few
decades. Analytic representation of the wind accretion for a perfect fluid onto a
Schwarzschild black hole has been found by \citet{Emilio1} and numerical treatments from Newtonian
and relativistic perspective were extensively studied by \citet{Donmez3, Donmez4, Cruz1, Cruz2}.

In the present paper, building onto the previous findings, we would like to study  the creation of the
shock cones and their dynamical
evolution in case of Bondi-Hoyle accretion in the background of the $4D$ EGB spherically
symmetric black hole in the strong gravitational region, numerically. Bondi-Hoyle accretion is an important
mechanism to explain Quasi-Periodic Oscillations (QPOs) observed by $X-ray$ telescope.
The companion star in the $X-$ray binary system provide the fast wind which causes to accreate
towards to the black hole\citep{Orosz1}. The accelerating single black holes due to the
other black holes could have Bondi-Hoyle accretion\citep{Lora1}.
Studying the QPOs on accretion disk in vicinity of a black hole is important to explore
the physical parameters of the black hole such as its spin and mass.
QPO arises in the inner accretion disk of the black hole binary and could be created
in terms of an oscillating, precessing hot flow in the truncated-disk geometry due to the
strong shocks. Therefore, it would be a further step to know whether the Gauss-Bonnet coupling
constant $\alpha$ can play a dominant role in the oscillation of the shock cone created during the
Bondi-Hoyle accretion. To reveal all these details, we numerically model the Bondi-Hoyle accretion
in the vicinity of $4D$ EGB black hole. We compute the shock cone structures for different values of
Gauss-Bonnet constant $\alpha$ and compare them with standard general relativistic solution, Schwarzschild
black hole. 

The plan of the paper is as follows: In Section \ref{Non-rotating Black Hole Solution of 4D EGB Gravity},
we briefly give summary of the recently proposed non-rotating black hole solution in $4D$ EGB gravity
and definition of the horizon of the black hole. In section \ref{GRHE},
we describe the conserved form of the general relativistic hydrodynamical equations, lapse function, and
shift vectors in $4D$ EGB black hole necessary in our numerical simulations. In Section \ref{BondiHoyle},
we  describe the theoretical framework of the Bondi-Hoyle
accretion for pressureless gas, and initial and boundary conditions used in our numerical simulation. The numerical
results and discussion are also given to show dependencies of  the disk dynamics, creation of shock cones,  accretion
rates, and mode power to Gauss-Bonnet coupling constant  $\alpha$. In Section \ref{Conclusion},  we  discuss the
implications from our numerical results  and draw the direction of future work.
Unless specified, we use geometrized units throughout
the paper for the speed of light and gravitational constant, $G=c=1$.


\section{Non-rotating Black Hole Solution of 4D EGB Gravity}
\label{Non-rotating Black Hole Solution of 4D EGB Gravity}
The theory of the static and spherically symmetric solution of the gravity can be driven by
summing Einstein-Hilbert action and higher order Lovelock invariants with the
vanishing bare cosmological constant \citet{Glavan1}.

\begin{eqnarray}
  S_{EH} + S_{GB} = \int{d^Dx\sqrt{-g}\left(\frac{M^{2}_{P}}{2}R + \alpha G\right)},
\label{EGB1}
\end{eqnarray}

\noindent
where $M_P$, $R$, $\alpha$, and  $G$  are the reduced Planck mass, the Ricci scalar of the space-time,
the Gauss-Bonnet coupling constant, and the Gauss-Bonnet invariant, respectively \citep{Glavan1, Cheng1}.
The theory of this black hole was already established for $D \geq 5$ \citep{Boulware1}. But the Gauss-Bonnet term
$G$ does not contribute to the $4D$ gravitational dynamics. The reason is that the equation of a motion contains
term $D-4$ and it will disappear while $D=4$. After rescaling the coupling constant
$\alpha \rightarrow \frac{\alpha}{D-4}$, the factor $D-4$ would be removed. So when the Lovelock theorem is
bypassed, it gives nontrivial dynamics, and the static and spherically symmetric black hole solution was
discovered\citep{Glavan1}.

The static and spherically symmetric black hole solution in four dimensional EGB gravity has the following
form

\begin{eqnarray}
 ds^2 = -f(r)dt^2 + \frac{1}{f(r)}dr^2 + r^2d\theta^2 + r^2sin(\theta)d\phi^2, 
\label{EGB2}
\end{eqnarray}

\noindent
where

\begin{eqnarray}
  f(r) = 1 + \frac{r^2}{2\alpha}\left(1 - \sqrt{1 + \frac{8 \alpha M}{r^3}} \right).
\label{EGB3}
\end{eqnarray}

\begin{figure}
  \center
  \vspace{0.6cm}
 \psfig{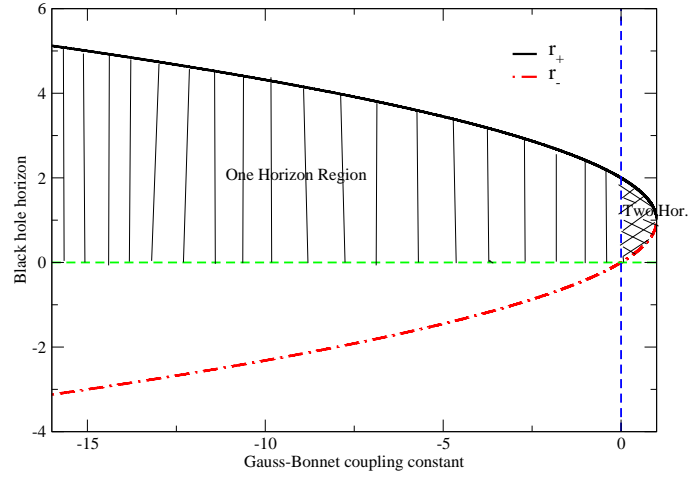}
 \caption{Black hole horizon radius $r_+$ and $r_-$ as a function
   of the Gauss-Bonnet coupling constant $\alpha$.}
\label{horizon}
\end{figure}

\noindent
Here $M$ is the mass of the black hole. The coupling constant parameter plays a dominant role defining
the black hole horizon. By solving $f(r)=0$, we can reach the following two horizons for non-rotating black hole
in $4D$ EGB gravity,

\begin{eqnarray}
  r_{\pm} = M \pm \sqrt{M^2-\alpha}, 
\label{EGB4}
\end{eqnarray}

\noindent
where, as seen in Fig.\ref{horizon}, the Gauss-Bonnet coupling constant is restricted to
$\alpha < 1$. If $\alpha > 0$, one has two horizons. One degenerate horizon is seen
when $\alpha = 0$. Otherwise, $\alpha < 0$, the black hole has only one horizon. It was belied that
static and spherically symmetric ansatz in Eq.\ref{EGB2} might not give the real
solution \citep{Glavan1} for a short radial distance $r^3 < -8 \alpha M$. As shown in  Fig.\ref{horizon},
the black hole in $4D$ EGB gravity exists when $\alpha < 0$. Therefore we would like to numerically model
Bondi-Hoyle accretion for possible values of Gauss-Bonnet coupling constant
either $0 <\alpha<1$ or $\alpha<0$ and  to compare them with
the theory and Schwarzschild solution.
As it  was also noted in
reference \citet{Guo1}, the Inner Stable Circular Orbit (ISCO) in the novel $4D$ Gauss-Bonnet
gravity can be greater or less than the one in Schwarzschild black hole solution depending on
the coupling constant $\alpha$. 

%



\section{General Relativistic Hydrodynamical Equations}
\label{GRHE}
General Relativistic Hydrodynamics (GRH) is a necessary theory to understand the dynamic
structures of the accretion disks around the black holes especially inside the region,
$r < 100M$ where the strong gravitational field is encountered. Here, $M$ is the mass of
the black hole. The covariant forms of the GRH equations are in the following form
\citep{Donmez1, Donmez2},

\begin{eqnarray}
 \bigtriangledown_{\mu}T^{\mu \nu}=0,  \;\;\;\;\;\;\;\;\;\;  \bigtriangledown_{\mu}J^{\mu}=0,
\label{GRH1}
\end{eqnarray}

\noindent
where $T^{\mu \nu}$ and $J^{\mu}$ are the stress-energy tensor and the matter current
density, respectively. The latin indices run from $1$ to $3$  and Greek indices run from
$0$ to $3$. 
In order to define the characteristic waves of the GRH equation, we simplify them 
neglecting the magnetic field and any type of the viscosity. Therefore the stress-energy
tensor  for a perfect fluid is,

\begin{eqnarray}
 T^{\mu \nu} = \rho h u^{\mu}u^{\nu} + P g^{\mu \nu},
\label{GRH2}
\end{eqnarray}

\noindent
where specific enthalpy $h = 1 + \epsilon + \frac{P}{\rho}$. $\rho$, $P$, $u^{\mu}$, and
$g^{\mu \nu}$ are the rest-mass density, ideal gas equation of state (pressure), four-velocity,
and four metric, respectively. The equation of state is $P = (\Gamma -1)\rho\epsilon$.
Here, $\Gamma$ is adiabatic
index which defines the compressibility of the fluid.

The form of GRH equations given in Eq.\ref{GRH1} is not suitable to use in High Resolution
Shock Capturing scheme (HRSC). Hence, GRH equations are written in conservation
form using $3+1$ formalism and we end up with the following first-order, flux-conservative
hyperbolic system,

\begin{eqnarray}
 \frac{\partial \vec{U}}{\partial t} + \frac{\partial \vec{F^i}}{\partial x^i} = \vec{S},
\label{GRH3}
\end{eqnarray}

\noindent
where $\vec{U}$, $\vec{F^i}$, and $\vec{S}$ are  conserved quantities, fluxes, and sources,
respectively.  Detailed formulations of these three vectors are given as the functions of
lapse function $\tilde{\alpha}$, shift vectors $\beta^i$, the determinant of the three-metric $\gamma$,
four-velocity $u^{\mu}$, three velocity $v_i$, Lorentz factor
$W = \tilde{\alpha}u^0=(1 - \gamma_{ij}v^iv^j)^{-1/2}$, rest-mass density $\rho$,
the components of the momentum vector $S_i$, pressure $P$, enthalpy $h$, and
the $4D$ Christoffel symbol $\Gamma^{\alpha}_{\mu \nu}$. The spectral decomposition of
the Jacobian matrix of the system $\partial \vec{F^i}/\partial \vec{U}$ is needed to use HRSC
scheme and they are given in \citet{Donmez1} with the full details.

The static and spherically symmetric solution of the black hole in $4D$ Gauss-Bonnet
gravity metric given in Eq.\ref{EGB2} is used to define the source term
appearing in the right hand side of Eq.\ref{GRH3}. The four-metric, three-metric, their
inverses, the $4D$ Christoffel symbol, and Lorentz factor can be easily driven
by having straightforward calculations. The lapse function for the EGB black hole is

\begin{eqnarray}
  \tilde{\alpha} = \left(1 + \frac{r^2}{2\alpha}\left(1 -
  \sqrt{1 + \frac{8 \alpha M}{r^3}} \right)\right)^{1/2},
\label{GRH4}
\end{eqnarray}

\noindent
and the shift vectors are,

\begin{eqnarray}
 \beta_r = 0, \;\;\;\;\;\;\;  \beta_{\phi} = 0, \;\;\;\;\;\;\; \beta_{\theta}= 0.
\label{GRH5}
\end{eqnarray}


\section{Bondi-Hoyle Accretion onto 4D EGB Black Hole}
\label{BondiHoyle}

\subsection{Theoretical Framework}
\label{Theoretical Framework}

\begin{figure}
  \center
   \vspace{0.3cm} 
 \psfig{file=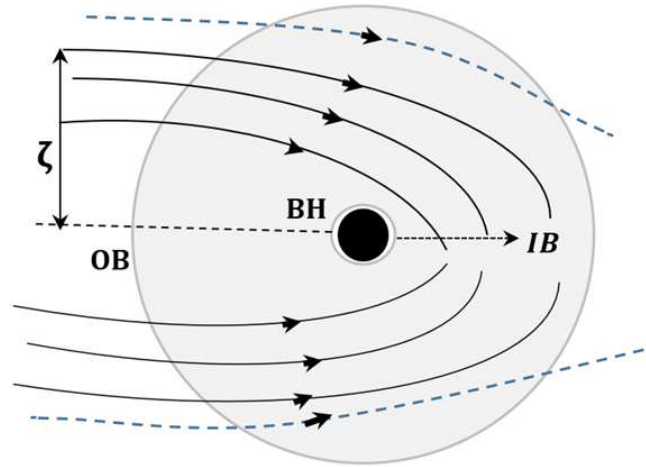,width=9.0cm}
 \caption{Artistic visualization of the Boyle Hoyle accretion around the black hole.
   BH $\equiv$ Black Hole, OB $\equiv$ Outer Boundary, and IB $\equiv$ Inner Boundary}
\label{BondiFig1}
\end{figure}

A homogeneous supersonic flow that has a relative velocity $v_{\infty}$ and density
$\rho_{\infty}$ at infinity is deflected by point mass $M$ due to the
warped space-time \citep{Hoyle1}. The Bondi-Hoyle accretion condition for particles is
defined with an impact parameter $\zeta$. As it is seen in Fig.\ref{BondiFig1},
if the pressure effect  is negligible,
the trajectory of the particle can be defined by conventional orbit. And then
accretion condition is,

\begin{eqnarray}
 \zeta < \zeta_{HL}=\frac{2M}{v_{\infty}}, 
\label{BonHoy1}
\end{eqnarray}

\noindent
where $\zeta_{HL}$ is commonly known as accretion radius or the stagnation point. The suggested
accreated matter could be computed by considering the trapped material inside the
gravitational potential. The accretion rate suggested by \citet{Hoyle1} is,

\begin{eqnarray}
  \dot{M}_{HL}= \pi\zeta^{2}_{HL}\rho_{\infty}v_{\infty} = \frac{4\pi M^2 \rho_{\infty}}{v^{3}_{\infty}}.  
\label{BonHoy2}
\end{eqnarray}

\noindent
The material falling towards the black hole that does not encounter to $\theta = 0$ axis
would not accreate  and it would escape from the system as seen in dashed (blue) lines in
Fig.\ref{BondiFig1}.

The theory given above does not exactly  describe the motion of the gas flow
accreated onto the black hole in the presence of the gas pressure in the
strong gravitation region but it helps us
define what the physical parameters are at infinity and how the
accretion mechanism works.

Bondi-Hoyle accretion is studied in the vicinity of the Schwarzschild and Kerr
black hole and called
spherically symmetric accretion. During the accretion, the Bondi radius would be
created and the accreated flow outside the radius is subsonic, and the disk
mass-density is almost uniform. But the gas inside the radius becomes supersonic and
it will be accreated towards to the black hole \citep{Donmez3, Donmez4}. The strong gravity
focuses the material behind the black hole (down-stream region) and the material would
be accreated. In this paper, we will numerically model the
Bondi-Hoyle accretion in the vicinity of
the static and spherically symmetric black hole in $4D$ Gauss-Bonnet gravity to reveal the
effect of Gauss-Bonnet coupling constant, not only on the shock cone structures but also on the
oscillation properties and QPOs of those shock cones.

\subsection{Initial and Boundary Conditions}
\label{Initial Conditions}

Numerical simulation of the Bondi-Hoyle accretion around static and spherically
symmetric non-rotating black hole defined in $4D$ EGB gravity gives us an opportunity
to understand how the dynamics of the accretion disk and shock cone would be effected by
the Gauss-Bonnet coupling constant $\alpha$. For this purpose, we numerically solve the
GRH equations
in spherical coordinate
on the equatorial plane $(r, \phi)$, i.e. $\theta = \pi/2$
(see Eq.\ref{GRH3})
assuming spherical symmetry
using the HRSC scheme based on  approximate Riemann
solver, further described in \citet{Donmez1}.

As explained in section \ref{GRHE}, we need
to know pressure to evolve GRH equation depending on given initial values. The pressure
is evaluated by using the perfect fluid equation of states $P = (\Gamma -1)\rho \epsilon$
with adiabatic index $\Gamma=4/3$. In order to perform the numerical simulation of the
Bondi-Hoyle accretion on the equatorial plane, the gas is injected from an outer boundary of the
upstream region with the following velocities,

\begin{eqnarray}
  V^r= \sqrt{\gamma^{rr}}V_{\infty}cos(\phi)   \;\;\;\;\;\;\;\;
  V^{\phi}= -\sqrt{\gamma^{\phi \phi}}V_{\infty}sin(\phi)
\label{IC1}
\end{eqnarray}

\noindent
$V_{\infty}$ is called the asymptotic velocity of the gas at infinity. The radial
and angular velocities of the gas injected from the outer boundary are also given
in terms of the components of the three-metric. The Eq.\ref{IC1} guarantees that
$V^2 = V_iV^i = V_{\infty}^2$ is valid everywhere along the computational domain.

The computational domain is defined on the equatorial plane with
    $r \in [r_{in}, 100M]$ and $\phi \in [0, 2\pi]$ in spherical coordinate. The uniform
    grid is used in the all models along the radial and the angular directions with $N_r=1024$ $\times$
    $N_{\phi}=512$ cells. So that the grid spacing is $(\triangle r, \triangle \phi)$ $=$
    $(4.45 \times 10^{-3} rad,9.4 \times 10^{-2}M)$ in geometrized unit. We used the dynamical
    time step size in order to satisfy the Courant-Friedrich-Lewy stability condition.
    In order to extract the 
    resolution dependencies of the numerical results,  we monitor the
    the mass-accretion rate result from one of our initial condition, changing grid
    resolution. We explore the results from three different resolutions,
    $N_r=512$ $\times$ $N_{\phi}=256$, 
    $N_r=1024$ $\times$ $N_{\phi}=512$, and $N_r=2048$ $\times$ $N_{\phi}=512$
    and found that the oscillation amplitude of the mass-accretion rate slightly
    decreases with increasing employed resolution. But the observed trend is true
    for all initial conditions. So that the important outcome of this paper,
    effects of Gauss-Bonnet coupling constant $\alpha$  in $4D$ EGB
    gravity on the accreated matter and shock cones created around the black hole,
    would not be effected from the resolutions.

In order to model the Bondi-Hoyle accretion numerically, a homogeneous supersonic flow
is injected from the upstream region $[\pi/2, 3\pi/2]$
at the location of the outer boundary using the
same analytic prescriptions given in Eq.\ref{IC1}. The computational domain is extended
from $r_{min}$, reported in Table \ref{table:Initial Models1} for different models, to
outer boundary $r_{max}=100M$ which is fixed for all models. $\phi$ goes from $0$ to $2\pi$.
The sound speed and the rest-mass density
are chosen as $c_{s,\infty}=0.1$ and $\rho_{\infty}=1$, respectively at the boundary.
And then the gas pressure is computed using the expression
$p = c_{s,\infty}^2\rho(\Gamma-1)/[\Gamma(\Gamma-1) - c_{s,\infty}^2\Gamma]$ accordingly
\citep{Donmez3}.  The asymptotic velocity is chosen as $V_{\infty} = 0.3$ with
the parameters mentioned above and given in Table
\ref{table:Initial Models1}.

\begin{table}
\footnotesize
\caption{The physical parameters and times: Gauss-Bonnet coupling
  constant $\alpha$, the inner boundary location of computational domain $r_{in}$,
  time to need to create a fully formed shock cone $t_{critial}$
  (required time for the steady state), and total
  simulation time  $t_{total}$. $M$ is the mass of the black hole.
 \label{table:Initial Models1}}
\begin{center}
  \begin{tabular}{cccc}
    \hline
    \hline
 $\alpha (M^2)$ & $r_{in} (M)$ & $t_{critical} (M)$ & $t_{total} (M)$ \\
 \hline
  $-12$  & $5$    & $\sim 1920$ & $19838$ \\
  $-7$    & $5$    & $\sim 1900$ & $25940$ \\
  $-5$    & $3.7$  & $\sim 1850$ & $21500$ \\
  $-0.8$   & $2.5$  & $\sim 1825$ & $17750$ \\
  $-0.01$   & $2.2$  & $\sim 1810$ & $16470$ \\
  $-0.0001$& $2.2$  & $\sim 1665$ & $21945$ \\
  $0.8$    &  $2.1$ & $\sim 1500$ & $18300$ \\
\hline
SCHW  & $2.2$  & $\sim 1650$ & $17000$ \\

\hline
\hline
  \end{tabular}
\end{center}
\end{table}

The initial injected values of a wind flow are used at upstream region
 at the outer boundary.  The continuation
is needed to avoid matter reflected back towards the black hole
at the outer boundary of the downstream region (from $3\pi/2$ to  $\pi/2$) along the radial coordinate.
It is treated simply by using the zeroth-order extrapolation for all primitive variables.
The inner boundary of computational domain along the radial distance
close to the black hole horizon
is implemented by using outflow boundary condition, handled simply copying first values
to ghost zones. Lastly, the periodic boundary condition is adopted along the $\phi$ direction.

\subsection{Numerical Results and Discussion}
\label{Numerical Results}

In order to reveal the effects of Gauss-Bonnet coupling constant $\alpha$
on the accreated material and the creation of shock cone, we first focus on
the injected matter from the upstream region of the computational domain. Later,
we wait some time to shock cone reaching the steady state and it is called $t_{critical}$.  The
critical times are varying from $1500 M$ to $1920 M$, shown in
Table.\ref{table:Initial Models1}. The evolution of these models are, at least, followed up
to $t=16470 M$ which is sufficient to extract oscillation properties of the shock
cone after reaches the critical time. The critical time is getting smaller for $\alpha > 0$
while it is increasing for  $\alpha < 0$. 

\begin{figure*}
  \vspace{1cm}  
  \center
  \psfig{file=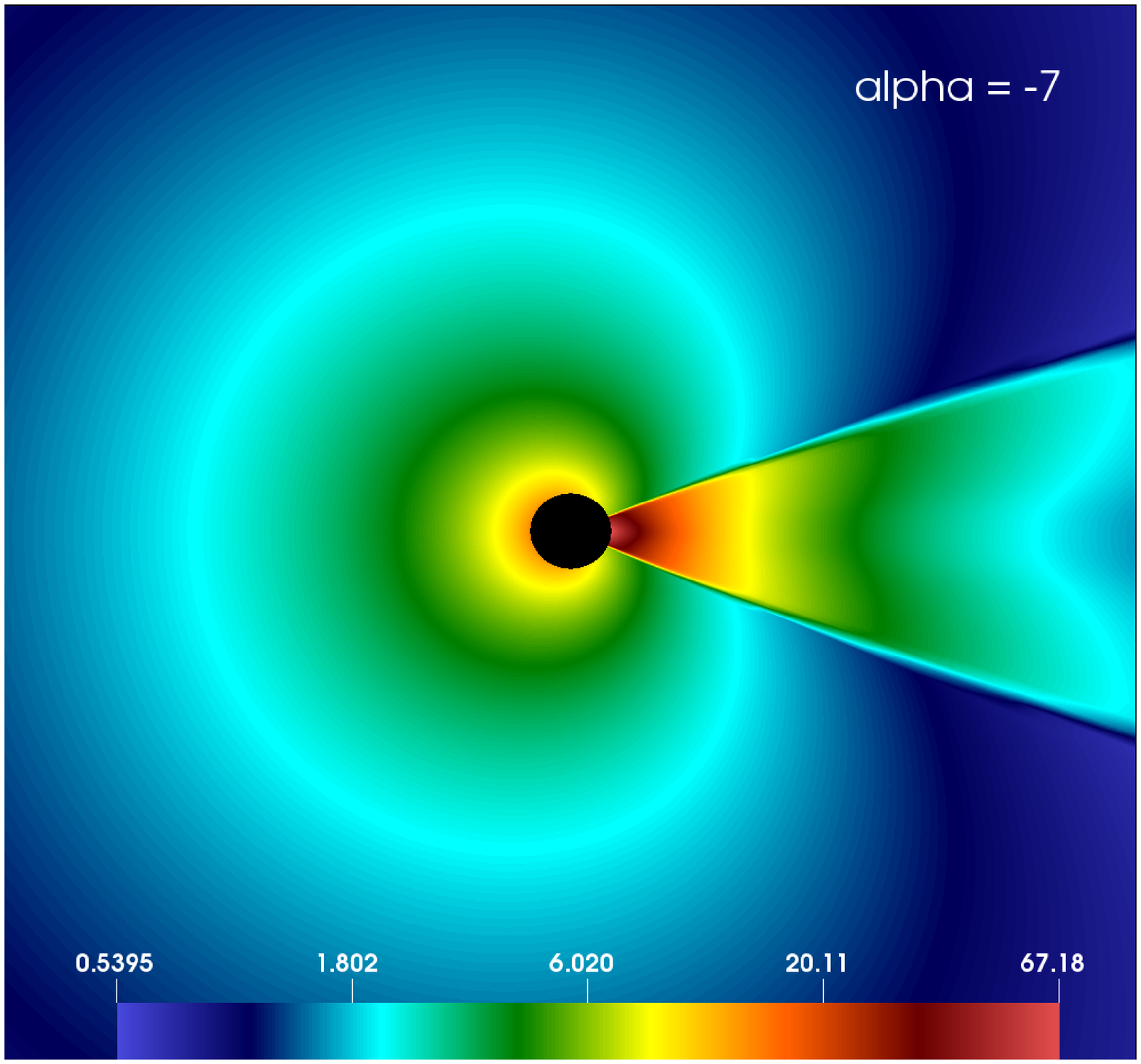,width=5.0cm}
  \psfig{file=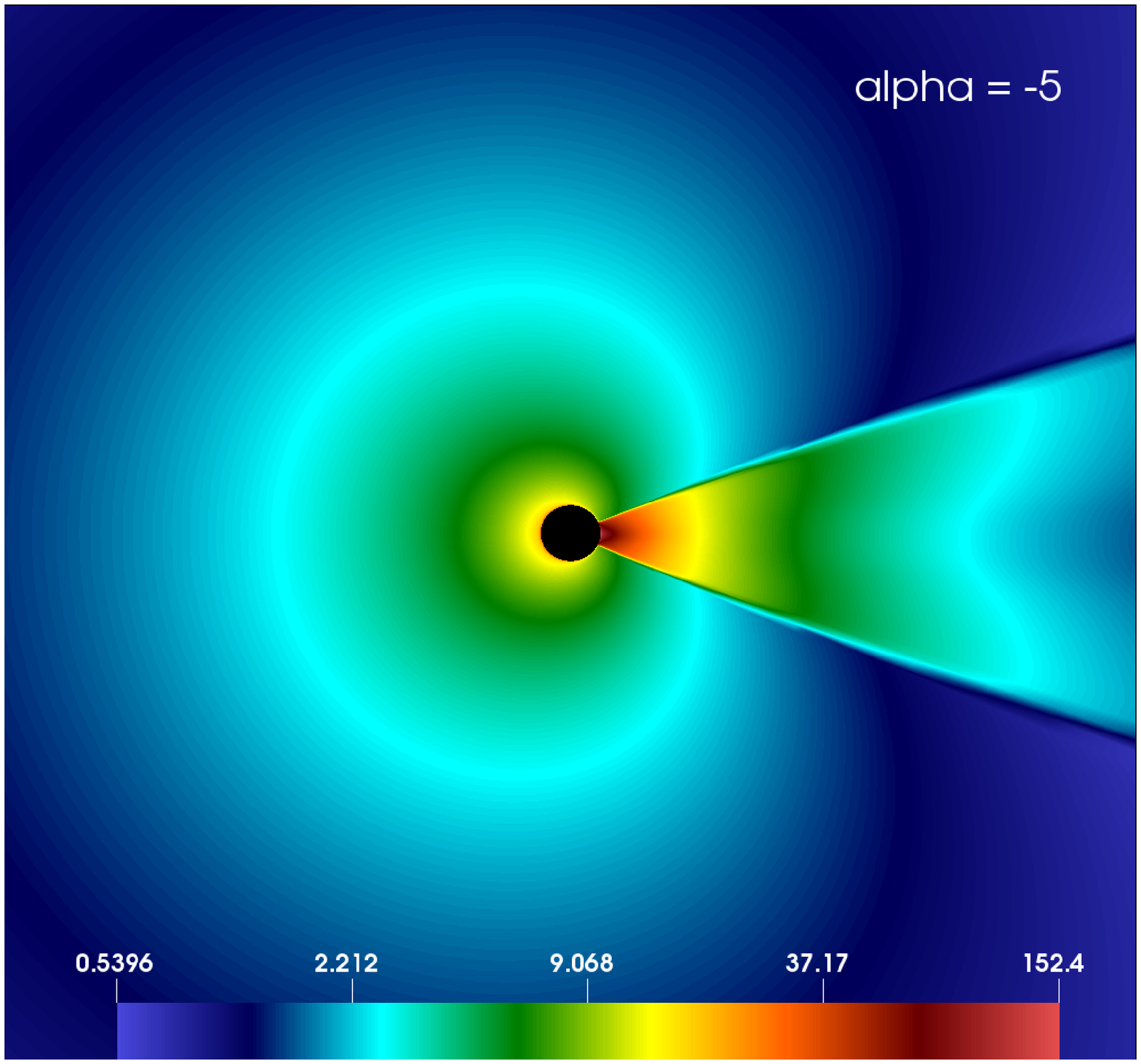,width=5.0cm}
  \psfig{file=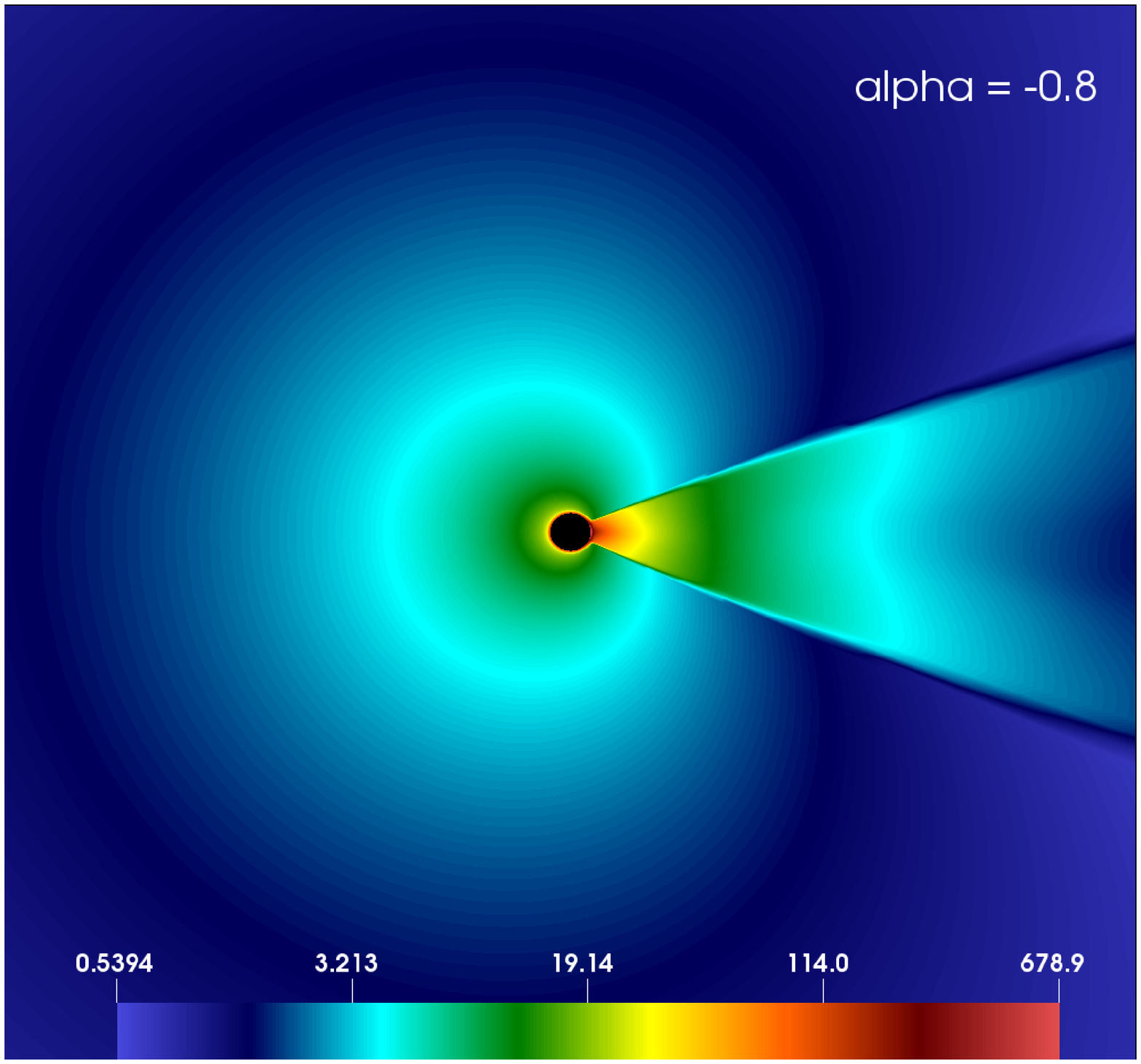,width=5.0cm}
  \psfig{file=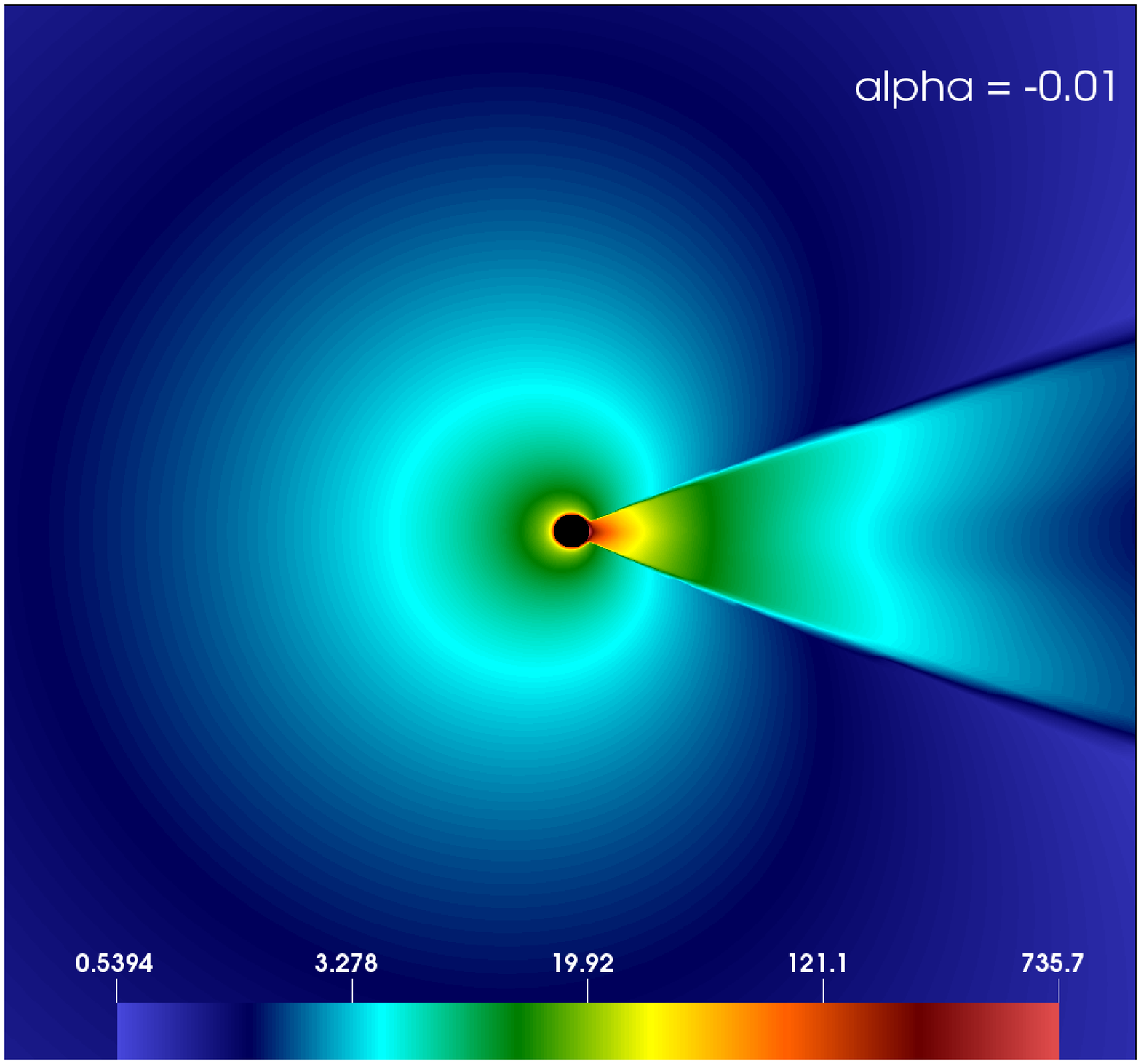,width=5.0cm}
  \psfig{file=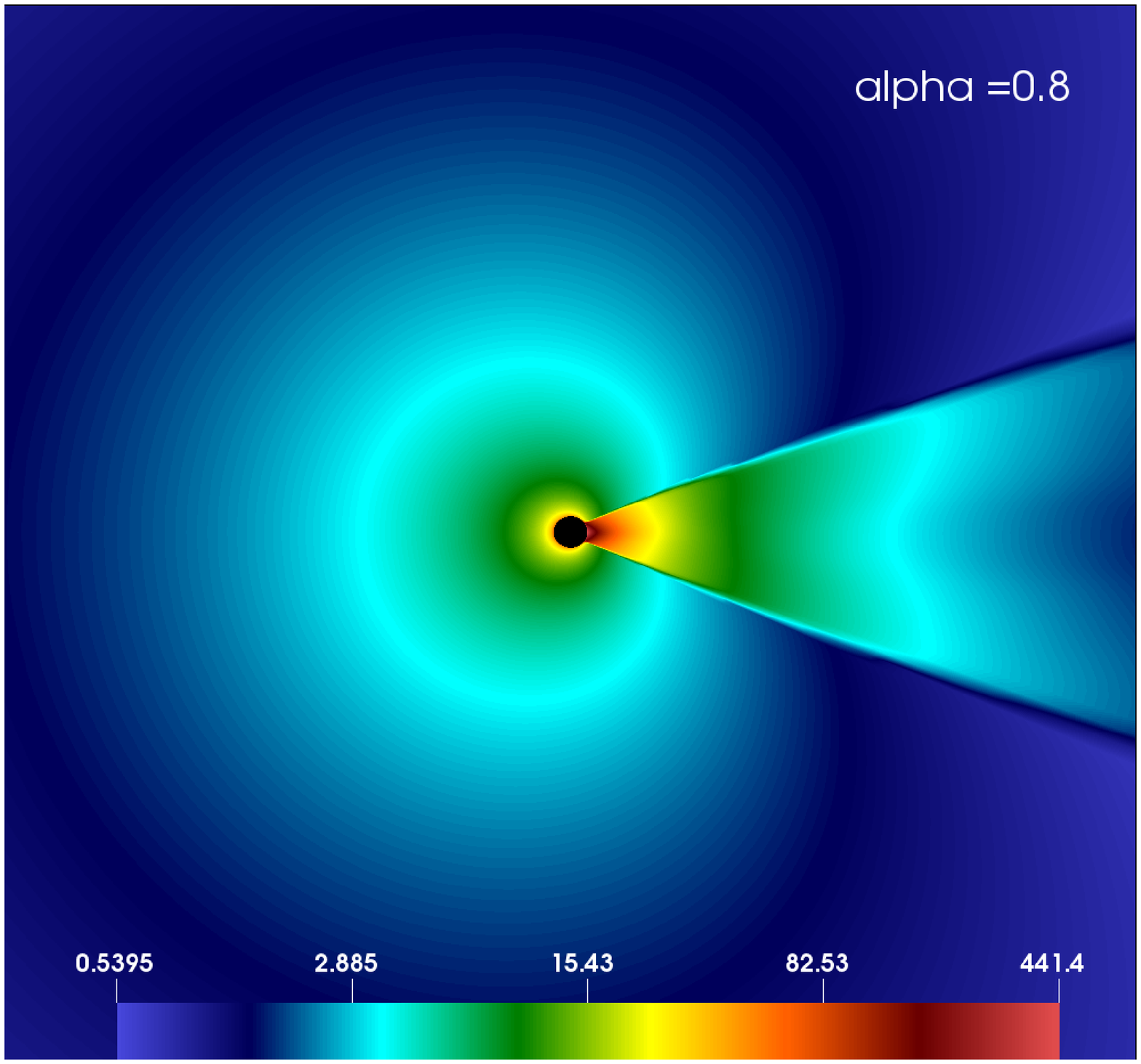,width=5.0cm}
   \psfig{file=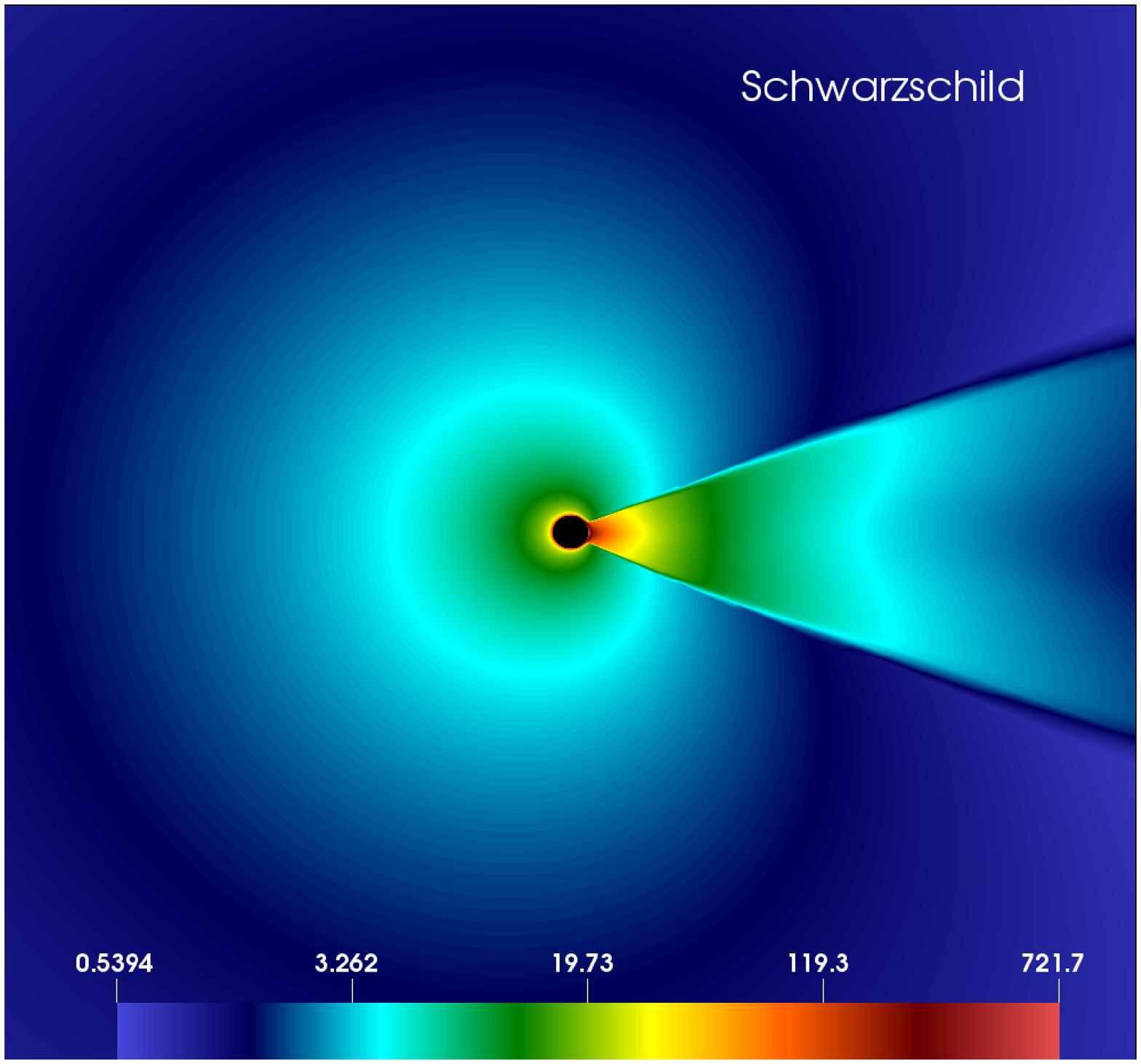,width=5.0cm} 
   \caption{Close-up view of the snapshots of the logarithmic rest-mass density on
     the equatorial plane at $t\sim15000 M$ much later than the shock cone reached 
     the steady state $t\sim1900 M$ for varying values of
     Gauss-Bonnet constant $\alpha$ and Schwarzschild black hole. The highlighted
   dynamic boundary is at $[x,y]$ $\rightarrow$ $[-70 M , 70 M]$.}
\label{Res1}
\end{figure*}

\begin{figure*}
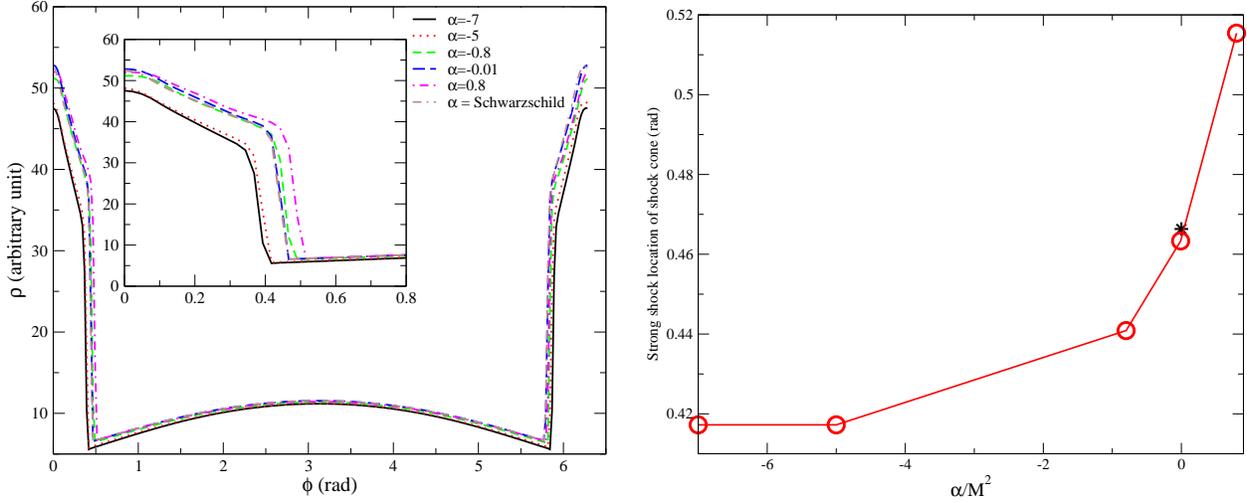

 \vspace{0.6cm}   
  \center
  \psfig{file=Fig61.eps,width=8.cm}
  \hfill
    \psfig{file=Fig62.eps,width=8.cm} 
 \caption{{\bf Left panel:} The rest-mass density as a function of the  angular coordinate $\phi$
   at $r=6.5M$ for the varying values of the Gauss-Bonnet coupling constant $\alpha$ and
   Schwarzschild black hole. The embedded plots indicates the left shock locations to
   see the differences for varying $\alpha$. {\bf Right panel:} The variation of the opening
   angle of the shock cone versus $\alpha$ computed at $\phi=\sim 0.5 rad$ in left panel.
   The black star represents the location for the Schwarzschild black hole. 
 }
\label{Res4}
\end{figure*}

The morphology of the accreated shock cones formed in downstream region are given  for
the various values of Gauss-Bonnet coupling constant $\alpha$, seen in Fig.\ref{Res1} where we
report the logarithmic rest-mass density of the accreated matter. The snapshots refer to $t\sim 15000 M$,
which is much later than the critical time needed to reach the steady state. The results
due to different $\alpha$ show a qualitatively similar behavior, although quantitative differences
appear in all models. The development of the shock cone
and its strong shock locations, also seen in Fig.\ref{Res4}, could cause a chaotic non-linear
phenomena inside the cone. The shock cones are attached to the numerical horizon, $r_{in}$ given
in Table \ref{table:Initial Models1} for different values of $\alpha$. It is know that
the certain amount of matter would pass across the cone and fall into the black hole.

Besides, understanding the dynamic evolution of an accretion disk due to the Bondi-Hoyle accretion and 
calculating the mass accretion rates allow us to find out more important features of  the shock
cone. The mass accretion rate is,

\begin{eqnarray}
 \frac{dM}{dt} = -\int_0^{2\pi}\tilde{\alpha}\sqrt{\gamma}\rho u^r d\phi,
\label{Mass_acc}
\end{eqnarray}

\noindent
where all the physical quantities are defined in Section \ref{GRHE}. 

\begin{figure}
  \center
 \psfig{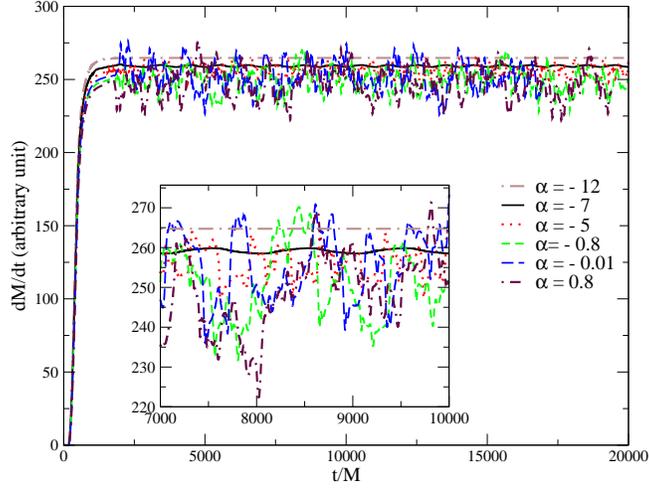}
 \caption{Mass accretion rate as a function of time for varying values of $\alpha$ at
   $r= 6.1 M$. The accretion rates reach the steady state around
   $t \simeq 1900 M$ in all models.}
\label{Res2}
\end{figure}

Here, we report the mass accretion rate at a fixed radial distance $r=6.1 M$
for various values of $\alpha$. The mass accretion  onto the $4D$
EGB black hole is clearly sensitive to the $\alpha$, as can be seen in Fig.\ref{Res2}.
It is clearly seen that increasing the value of 
$\alpha$ towards  zero leads to more violent phenomena inside the shock cone and thus
creates severe oscillations after the shock cone reaches the steady state.
The high amplitude oscillation is an indicator of instability fully developed during the evolution.
We have also
confidence that the inner boundary of the computational domain does not play any role on oscillation
properties of the shock cone. The oscillating properties of the mass accretion rates are not the same
although the inner boundaries for $\alpha = -12$ and $\alpha = -7$ are in the same locations,
On the other hand, oscillation inside the shock cone is considerably dissipated by the larger
values of the negative $\alpha$.

The occurrence of the instability is the result of
    moving wave-like  stretching  into the medium which has a lower density. 
    Overall, it is fair to say that, the
    instability inside the shock cone has been confirmed numerically, but
    it is still in debate whether the physical mechanisms driving the instabilities.
    The instability could be depended on the physical nature of the wind, such as
    sound speed, matter velocity, and Bondi accretion radius.
    A  very  detailed analysis of the unstable behavior of Bondi-Hoyle
    accretion flows was investigate by \citet{Foglizzo}, suggesting that the
    instability might be of adjective-acoustic nature  in  the  case  of
    shocks cones around the black hole.

\begin{figure}
  \center
 \psfig{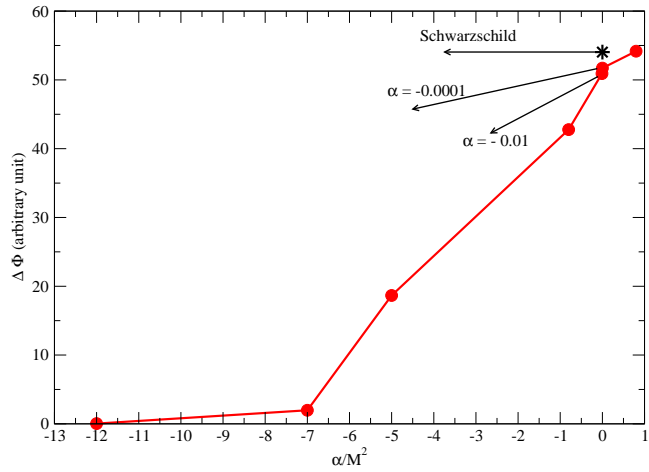}
 \caption{The behavior of the maximum oscillation amplitude $(\Delta \Phi)$ of the
   shock cone with varying
   Gauss-Bonnet coupling constant $\alpha$ after the shock cone is fully formed.
   Star seen in figure
 represents the oscillation amplitude of the shock cone around the Schwarzschild black hole.}
\label{Res3}
\end{figure}

After showing the behavior of the accretion rates in vicinity of $4D$ EGB black hole, we can now
discuss the maximum value of oscillation strength using the mass accretion rate.
Fig.\ref{Res3} represents the maximum oscillation amplitude $(\Delta \Phi)$ of
the shock cone with various Gauss-Bonnet coupling constant $\alpha$  after
the shock cone is fully formed. This is clear evidence  that the significant oscillations happen
when $\alpha$ is getting closer to zero as compared to high negative values.
Together with the dependency of the accretion rate to  $\alpha$, we have also investigated
the convergence  of EGB gravity solution to the general relativistic one.  As it can be seen
in Fig.\ref{Res3}, $(\Delta \Phi)$ around the  $4D$ EGB black hole converges to Schwarzschild one
when $\alpha \rightarrow 0$.

To have deeper understanding  of Fig.\ref{Res1}, we extract the information  along
the angular direction $(\phi)$ at a fixed radial coordinate $r=6.5M$. The left panel of
Fig.\ref{Res4} shows one-dimensional profiles of the rest-mass density for different
values of Gauss-Bonnet coupling constant $\alpha$ and Schwarzschild solution. The strong
shocks are created at the border of the shock cone. Therefore, a sharp transition is seen
in density. The sharp transition location has a lowest density throughout the disk since
along this location gas falls towards to the black hole along the streamlines, supersonically. 
The locations of streamlines seen in left side of the left panel of Fig.\ref{Res4} versus $\alpha$
are given in the right panel of the same figure. Increasing in $\alpha$ (going from $-7$ to $0.8)$
causes a slight exponential increase in the location of streamline, seen in the right panel of Fig.\ref{Res4}.
Hence the shock opening angle appeared in the downstream  side of the computational domain expands along
the angular direction.

In order to uncover the effects of Gauss-Bonnet coupling constant $\alpha$ on the
instability created during  evolution and especially
inside the shock cone after it reaches  the steady state, we numerically study the Fourier
mode analyze to compute  the saturation point and to analysis the characterization of the
instability. The Fourier mode $m=1$ is performed and the growth rates are obtained for the
the rest-mass density power using the following equation \citep{Donmez5},

\begin{eqnarray}
 P_m = \frac{1}{r_{out}-r_{in}} \int_{r_{in}}^{r_{out}}{ln ([Re(w_m(r))]^2+[Im(w_m(r))]^2) dr},
\label{Mod1}
\end{eqnarray}

\noindent
where $r_{out}$ and $r_{in}$ are outer and inner boundary of the computational domain,
respectively. The real and imaginary parts are
$Re(w_m(r)) = \int_{0}^{2 \pi}\rho (r,\phi) cos(m \phi) d\phi$ and
$Im(w_m(r)) = \int_{0}^{2 \pi}\rho (r,\phi) sin(m \phi) d\phi$.  

\begin{figure}
\vspace{0.6cm}   
  \center
 \psfig{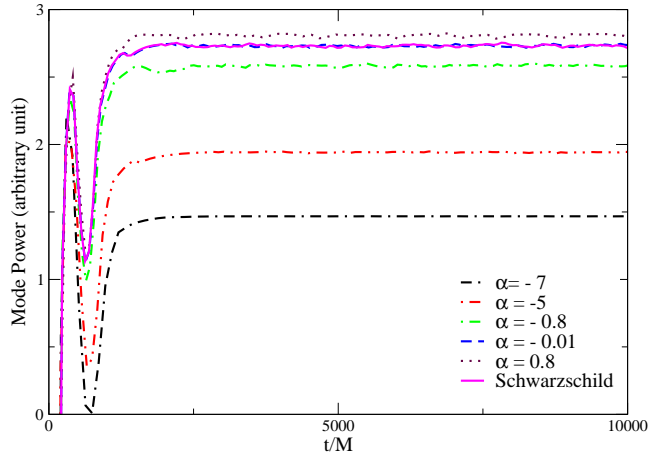}
 \caption{$m=1$ mode growth evolution around the EGB black hole for different
   values of $\alpha$ and Schwarzschild
   black hole. The region of interested is zoomed to see the saturation points and oscillations
 of the mode powers after they reached the steady state.}
\label{mode_power}
\end{figure}

As seen in Fig.\ref{mode_power}, the instability mode grows in the beginning of the
simulation until $t\sim 304 M$ for $\alpha = -7$
but the time is slightly getting greater when $\alpha$ converges to $0$. And then we
have witnessed decreasing in the mode for a short time scale $(t\sim200M)$. Later,
the $m=1$ mode creates an exponential growth again until $t\sim1400M$. The $m=1$ mode
power reaches a maximum
for  the $\alpha$ getting close to zero and $0< \alpha < 1$, although the modes in all models saturate
almost at the same time $t\sim1400M$. On the other hand, both  $\alpha$ close to zero and
Schwarzschild black hole power modes exhibit the same behavior during the evolution.

\begin{figure}
\vspace{0.6cm}   
  \center
 \psfig{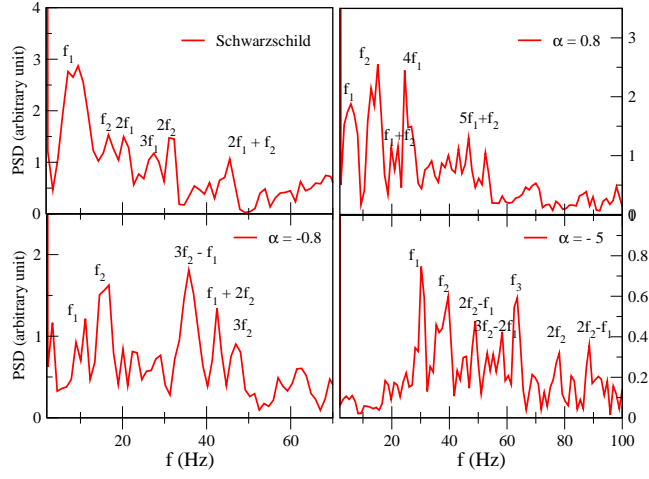}
 \caption{Power spectra of the mass accretion rate for different values of
   Gauss-Bonnet coupling constant $\alpha$ and Schwarzschild solution computed at $r=6.1 M$.
 The mass of the black hole was assumed to be $M = 10 M_{\odot}$.}
\label{PSD}
\end{figure}

Fourier analysis of the computed data allows us to extract a lot of information
about emerged phenomenology and
their connections with Gauss-Bonnet coupling constant $\alpha$. In Fig.\ref{PSD},
for instance, we obtain the power spectra using the rest-mass accretion rate computed during
the time evolution for different values of $\alpha$ and Schwarzschild black hole.
We have  also conducted an extra test to find the dependency of power spectra to
radial position and found that the mode of oscillation does not depend on the radial position.
It means that the modes are global eigenmodes of the oscillating shock cone. As it is seen in
Fig.\ref{PSD}, there are two genuine eigenmodes appearing in the power spectra and the rest are the
results of nonlinear couplings of these genuine modes. For Bondi-Hoyle accretion around the
Schwarzschild solution, $f_1 = 9.5 Hz$ and $f_2 = 17 Hz$ are genuine eigenmodes, while
$20 Hz$, $27.7 Hz$, $32 Hz$, and $45.5 Hz$ are the nonlinear couplings of those genuine eigenmodes
with a $2 Hz$ of error bar.  These nonlinear couplings are $2f_1$, $3f_1$,  $2f_2$, and
$2f_1 + f_2$, respectively. Similarly, for  $\alpha = 0.8$,
$f_1 = 6 Hz$ and $f_2 = 13.5 Hz$ are genuine eigenmodes, while
$21 Hz$, $24 Hz$, and  $46.5 Hz$ are the nonlinear couplings of those genuine eigenmodes
with a $2 Hz$ of error bar.  These nonlinear couplings are $f_1+f_2$, $4f_1$, and  $5f_1+f_2$,
respectively. For  $\alpha = -0.8$,
$f_1 =11 Hz$ and $f_2 = 15.5 Hz$ are genuine eigenmodes, while
$36 Hz$, $42 Hz$, and  $47.5 Hz$ are the nonlinear couplings of those genuine eigenmodes
with a $2 Hz$ of error bar.  These nonlinear couplings are $3f_2 -f_1$, $f_1+2f_2$, and  $3f_2$,
respectively. For  $\alpha = -5$,
$f_1 =30 Hz$, $f_2 = 39.5 Hz$, and $f_3 = 63 Hz$ are genuine eigenmodes, while
$49 Hz$, $58.3 Hz$,  $78 Hz$, and $89 Hz$  are the nonlinear couplings of those genuine eigenmodes
with a $2 Hz$ of error bar.  These nonlinear couplings are $2f_2 -f_1$, $3f_2-2f_1$,  $2f_2$,
and $3f_2-f_1$, respectively. This behavior is  expected behavior of the  nonlinear equations in
the physical system in the limit of small oscillations \citep{Landau1}.
It is also important to note that the genuine modes and their nonlinear couplings show
different behavior for $\alpha = -5$. There are three main differences between $\alpha$ close
to zero (Schwarzschild solution) and  negative value of $\alpha$ when $\alpha = -5$. First, there are
two genuine modes for $-0.8 \leq \alpha \leq 0.8$ while there are three for $\alpha = -5$.
Second, the frequency of the first genuine mode $f_1$ is getting greater when $\alpha$ goes from
$0.8$ to $-5$.
Third, the frequencies of these genuine eigenmodes are much higher in $\alpha = -5$ than 
the other three cases. As it is expected, the amplitude of genuine modes are getting
smaller when $\alpha $ is getting larger in negative direction.

%


\subsection{Comparison of 4D EGB and Schwarzschild  Black Holes}
\label{Comparision of 4D EGB and Schwarzschild  Black Holes}

\begin{figure}
\vspace{0.6cm}   
  \center
 \psfig{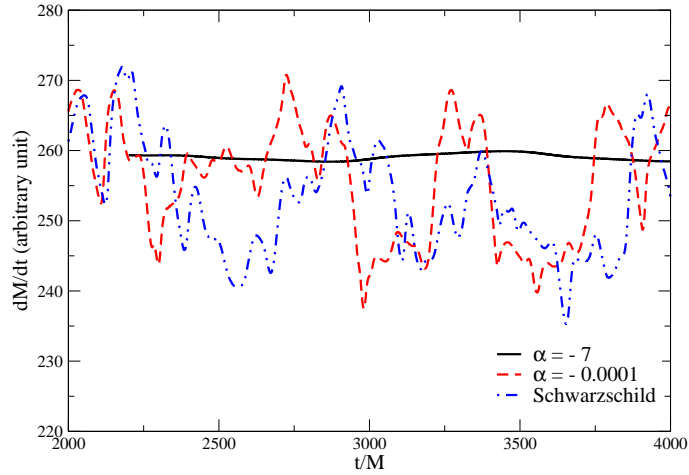}
 \caption{Comparison of the mass accretion rate for $\alpha =-7$ and $-0.0001$ with
   Schwarzschild solution as a function of time. The mass accretion rates are plotted after
 the shock cone reaches to the steady state.}
\label{compare1}
\end{figure}

To illustrate the consistency of the results found from the static spherically symmetric
black hole solution in a novel $4D$ EGB gravity with the Schwarzschild one, the Schwarzschild
solution of the Bondi-Hoyle accretion using the same initial conditions is performed. 
Fig.\ref{compare1} shows the comparison of the accretion rates between EGB and Schwarzschild
solutions using certain values of the Gauss-Bonnet coupling constant $\alpha$. The accretion rates
are plotted after the initial phase of relaxation of the shock cone. As it is also seen in
the previous section, the computed accretion rate for
the $\alpha =-0.0001$ shows similar behavior (oscillation + amplitude) with the Schwarzschild solution.
Obviously, tendency of $\alpha$ to  reach zero  would give the solution in the  general relativity. 
But decreasing in the parameter $\alpha$ causes a significant change in the oscillation properties of the
accreated disk. So that a smaller $\alpha$ would cause to form a less luminous,  cooler, and less efficient
shock cone around the black hole. This is in agreement with a thin accretion solution around
the $4D$ EGB black hole \citep{Cheng1}.

\begin{figure}
\vspace{0.6cm}   
  \center
 \psfig{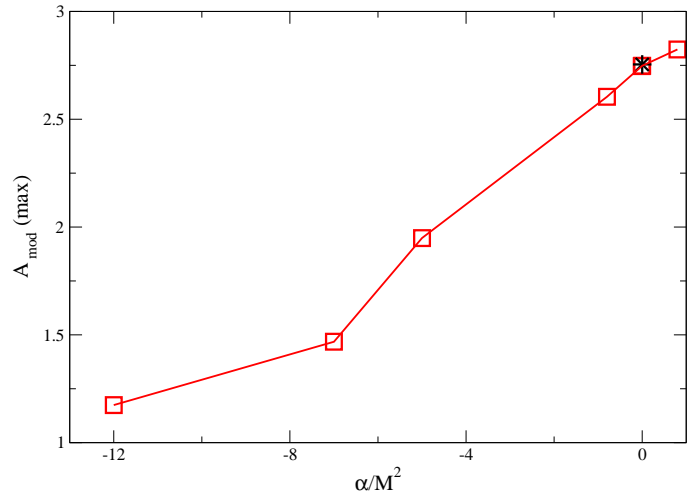}
 \caption{The maximum value of the $m=1$ power mode $(A_{mod}(max))$ versus
   Gauss-Bonnet coupling constant $\alpha$. Each value of power mode is extracted from the data
   after growth rate reaches to the saturation point. The star symbol
 represents the maximum power mode for the Schwarzschild black hole.}
\label{compare2}
\end{figure}

Power spectrum data has a clear signature about the creation of shock cone instability  for various values
of Gauss-Bonnet coupling constant  $\alpha$ and Schwarzschild black hole.
Fig.\ref{compare2} represents how the maximum value of the $m=1$ power mode $(A_{mod}(max))$ changes
with $\alpha$ and Schwarzschild black hole. $A_{mod}(max)$ increases with an increasing $\alpha$ and
it goes to the Schwarzschild solution when $\alpha \rightarrow 0$ (e.g. see the star
and the square symbols at $\alpha \sim 0$ in Fig.\ref{compare2}). For the positive values of $\alpha$,
it is seen in Fig.\ref{compare2} that increasing in $\alpha$ produces higher values of the maximum  power
spectrum.

%


\section{Conclusion}
\label{Conclusion}

We have performed  a systematic investigation of Gauss-Bonnet coupling constant $\alpha$
during the Bondi-Hoyle accretion onto a non-rotating black hole in $4D$ EGB gravity. The effect
of  $\alpha$ on the shock cone and its oscillation properties have been extensively studied
when $-12 \leq \alpha < 1$. The numerical simulation of Bondi-accretion onto Schwarzschild black hole
using the same initial condition is also performed to compare the general relativistic solution with
$4D$ EGB gravity one.

We investigate how the physical features of the shock cone and its oscillation
properties depend on $\alpha$ and find the following outcomes summarized
in Table \ref {table:Summarry}.

\begin{itemize}
\item We have found that the Bondi-Hoyle accretion around $4D$ EGB black hole with a
negative Gauss-Bonnet coupling constant produces physical solution
even it is very close to the black hole horizon. The same was also confirmed
by \citet{Guo1}  for the geodesic motions of time-like and null particles. 
\item The numerical simulations reveal that  moving matter from upstream side of the computational
domain creates a steady state shock cone in a short time scale. The required time having the shock cone
in downstream region is called critical time which decreases for $\alpha > 0$
while increases for  $\alpha < 0$.
\item The key indicator of having instability inside the shock cone is  a high
amplitude oscillation.  The increase in the $\alpha$
that is getting close to zero, would lead to  more violent phenomena inside the shock cone.
Therefore the severe oscillations are created after the shock cone reaches the steady state.
In addition, the oscillation inside the shock cone is considerably dissipated by the larger
values of the negative $\alpha$. 
\item The shock opening angle slightly depends on the variation of $\alpha$. Going from $\alpha=-7$ to
$\alpha=0.8$ would lead an exponential growth in the location of the shock cone. 
\item The power mode $m=1$  reaches a saturation point almost at the same time for all models.
The $m=1$ mode power gets  the maximum value when $\alpha$ is  close to zero and $0< \alpha < 1$.
In addition, the power modes calculations for the Schwarzschild black hole and for $\alpha$ close to zero
show the same behavior during the evolution.

\end{itemize}

\begin{table}
\footnotesize
\caption{ The effects of Gauss-Bonnet coupling constant $\alpha$
  to the accretion mechanism, shock cones, and physical  interpretation
  are summarized. $D$ represents decreasing while  $I$
  represents increasing. $SS$, $SC$, $PD$ stand for Steady State, Shock Cone , and
  Power Mode, respectively.
 \label{table:Summarry}}
\begin{center}
  \begin{tabular}{ccccc}
    \hline
    \hline
    & & Time to reach the SS  & &  \\
   $\alpha = -12$ & $\Longleftarrow (I)$ & $\alpha = 0$ & $\Longleftarrow (I)$ & $\alpha < 1$ \\
\hline 
 & &  Density  inside the SC  & & \\
   $\alpha = -12$ & $\Longleftarrow (D)$ & $\alpha = 0$ & $\Longleftarrow (I)$ & $\alpha < 1$ \\
\hline 
  & & Max. PD after saturation  & &      \\
$\alpha = -12$ & $\Longleftarrow (D)$ & $\alpha = 0$ & $\Longleftarrow (D)$ & $\alpha < 1$ \\
\hline 
  & & The shock location  & &      \\
$\alpha = -12$ & $\Longleftarrow (D)$ & $\alpha = 0$ & $\Longleftarrow (D)$ & $\alpha < 1$ \\

\hline
\hline
  \end{tabular}
\end{center}
\end{table}

In addition to recovering many important features of the shock cones for varying $\alpha$,
the novel features of the shock cones which traps the pressure modes inside the
high density are also extracted by using Fourier transform. QPOs are  computed for
the Schwarzschild  and different values of $\alpha$, i.e. $\alpha=0.8$, $\alpha=-0.8$, and
$\alpha=-5$. The global genuine modes and their nonlinear oscillation counterparts
are obtained in all cases. $\alpha$ influences not only the amplitude of modes but also
their absolute frequencies of the trapped modes. There are three genuine modes found
in $\alpha=-5$ while it is  two for $\alpha=0.8$ and $\alpha=-0.8$ including  the Schwarzschild one.
It is also important to note that the higher absolute frequencies are found in case of 
$\alpha=-5$. As a result, having three genuine modes need to be confirmed by doing more
accurate simulations in the negative  $\alpha$ direction.

Possible applications of the numerical results discussed here could
    be Sagittarius $A^*$ (Sgr $A^*$) and High Mass $X-$ray Binaries (HMXBs). It is possible
    to have a direct comparison between numerically computed QPOs and QPOs observed in
    the $X-$ray spectra of these sources.  Sgr $A^*$ black hole is located at the center of
    our own Galaxy and believed to be a supermassive black hole with a mass
    $4.6 \pm 0.7 \times 10^6 M_{\sun}$ which is the one of target of Event Horizon Telescope
    (EHT)  \citep{Akiyama1, Akiyama2, Akiyama3, Abbott1, Abbott2}. HMXBs are composed of
    black hole and of an OB star. We are planing to do more direct comparisons in the
    forthcoming  paper. In this paper, we will investigate the effects  of the black hole
    rotation parameter and Gauss-Bonnet
    coupling constant $\alpha$ on the shock cone dynamics 
    around the rotating  EGB black hole.

Finally, we have carried out a comparison of $4D$ EGB solution with Schwarzschild one. 
Obviously, it is seen from our numerical simulations that the tendency of $\alpha$ to reach zero
would yield solution in general relativity.


\section*{Acknowledgments}
The author thanks to the anonymous referee for constructive comments on the original manuscript.
All simulations were performed using the Phoenix  High
Performance Computing facility at the American University of the Middle East
(AUM), Kuwait.\\

\end{document}